# Effects of Pb doping on the Thermoelectric Properties of $Tl_{8.67}Pb_xSb_{1.33-x}Te_6$ Materials

Wiqar Hussain Shah, Aqeel Khan, Waqar Adil Syed
Department of Physics, Faculty of Basic and Applied Sciences, International Islamic University,
H-10, Islamabad, Pakistan

**Abstract:**

We present the effects of Pb doping on the thermoelectric properties of Tellurium Telluride $Tl_{8.67}Pb_xSb_{1.33-x}Te_6$ (x=0.61, 0.63, 0.65, 0.67, 0.68, 0.70), prepared by solid state reactions in an evacuated sealed silica tubes. Additionally crystal structure data were used to model the data and support the findings. Structurally, all these compounds were found to be phase pure as confirmed by the XRD and EDX analysis. The Seebeck co-efficient ($S$) was measured for all these compounds which show that $S$ increases with increasing temperature from 295 K to 550 K. The Seebeck coefficient is positive for the whole temperature range, showing p-type semiconductor characteristics. Complex behavior of Seebeck coefficient for Pb doped compounds has been observed that at room temperature, the values of S for *Pb* based compounds have complex behavior, first S decreasing with increase in *Pb* concentration i.e. x=0.65, and then S increases with increase in *Pb* contents upto x=0.70. Similarly the electrical conductivity ($\sigma$) and the power factors have also complex behavior with *Pb* concentrations. The power factor ($PF=S^2\sigma$) observed for $Tl_{8.67}Pb_xSb_{1.33-x}Te_6$ compounds are increases with increase in the whole temperature range (290 K-550 K) studied here. Telluride's are narrow band-gap semiconductors, with all elements in common oxidation states, according to $(Tl^+)_9(Sb^{3+})(Te^{2-})_6$. Phases range were investigated and determined with different concentration of *Pb* with consequents effects on electrical and thermal properties.

Corresponding author: wiqarhussain@yahoo.com

# I. Introduction:

Thermoelectrics (TE's), as one of the most promising approaches for solid-state energy conversion between heat and electricity, is becoming increasingly important within the last decade as the availability and negative impact of fossil fuels draw increasing attention. Various TE's materials with a wide working temperature range (from 10 to 1000 K) for different applications in cooling and power generation have been extensively studied [**1-5**]. Tellurium telluride (($Tl_5Te_3$)) is one of the most studied and used intermediate temperature TE materials with good thermoelectric properties , suitable for power generation applications such as waste heat recovery [**1**] and potentially in solar energy conversion [**2**]. Tellurium telluride based alloys are very attractive thermoelectric (TE) materials due to their high energy conversion efficiency at ambient temperature without requiring any driving parts or cooling system in electronic devices [**3**]. Many contributions in the vast field of thermoelectrics focus on the development and characterization of advanced thermoelectric materials [**4-5**]. Thermoelectric power generation is foreseen to play a much larger role in the near future due to the need for alternative source of energies because of declining natural resources as well as the increasing efficiency of thermoelectric materials. The later is the consequences of the discoveries of new materials as well as of improvements in established materials by, for example nano-structuring or band structure engineering. Thermoelectric materials can convert heat into electricity not only to harvest electrical energy from waste heat of different sources, such as in automotives [**6**], but also to enhance the output of solar-thermal-electric energy convertors [**7**].

The effectiveness of a material for thermoelectric applications is determined by the dimensionless figure of merit, $ZT=S^2\sigma T/\kappa$, where $S$ is the Seebeck coefficient, $\sigma$ is the electrical conductivity, $\kappa$ is the thermal conductivity, and $T$ is the absolute temperature [**8**]. The electrical properties are determined by the power factor, defined as $S^2\sigma$. The power factor can be optimized as a function of the carrier concentration; therefore, the thermal conductivity must be reduced to obtain maximum *ZT*. The conversion efficiency of a materials increases with increasing thermoelectric figure of merit. For large scale applications, the conversion efficiency and thus the figure of merit needs to be enhanced for this technology to become economically viable.

We have explored materials with more complex compositions and structures which is likely to have more complex electronic structures that could give rise to high TE performance. Good TE materials require an unusual combination of electrical and thermal properties. The challenge lies in achieving simultaneously high electrical conductivity $\sigma$, high TE power *S*, and low thermal conductivity k. All three of these properties are determined by the details of the electronic structure (band gap, band shape, and band degeneracy near the Fermi level, electronic concentrations) and scattering of charge carriers (electrons or holes) and thus are not independent from each other.

Attempts over more than 30 years period to improve thermoelectric efficiency of state of the art materials have met with very limited success, and solid solution based on $Bi_2Te$, and $TlSbTe_6$ remains the best thermoelectric materials for room and mid-temperature operation [9]. Interests are currently focused on the search for novel and more efficient thermoelectric materials for cooling and power generations. As mentioned, thermoelectric efficiency of a material is characterized by figure-of-merit ZT, traditionally, high ZT values are found in moderate to heavily doped semi-conductors system. In semiconductors, the thermal conductivity is dominated by the lattice component thus the problem of producing a high ZT value reduces to maximising the power factor $S^2 \sigma$ and minimizing $\kappa_L$. The power factor depends substantially on details of the electronic band structure and carrier scattering mechanism. The important condition for achieving a large power factor is by maximizing the parameter $S^2 \sigma$. For most of the materials the requirement of high Seebeck co-efficient and high electrical conductivity are mutually incompatible. These conditions can be achieved in materials, having complex band structure, high degree of degeneracy with several co-existing bonding types and scattering mechanism. A group of materials which satisfy these requirements are the *Pb* doped ternary compounds of $Tl_{8.67}Sb_{1.33}Te_6$. There is very limited information on the thermoelectric properties of *Pb* doped $Tl_{8.67}Sb_{1.33}Te_6$ compound. Structural, electrical and thermoelectric properties of these compounds have been studied over temperature rang 300 to 550 K as a function of deviation from stoichiometry.

Most efficient thermoelectric materials known to date are semi-conducting heavy-metal chelcogenides and antimonides. To optimize the properties, these materials are typically doped to get charge carrier concentrations of the order of $10^{19}$-$10^{21}$ carriers per cubic centimeter [10-12]. Heavy elements are known to contribute to low thermal conductivity an important asset of enhanced thermoelectric properties. This particularly appears to be true for materials containing thallium [13-15]. Various thallium chalcogenides bestow with an extremely low lattice thermal conductivity with values significantly below 1 W/mK, likely a consequence of the high mass of thallium and its tendency to complex low symmetry coordination spheres [16]. The crystal structure type adopted by $Tl_9Sb_1Te_6$ is an ordered variant of $Tl_5Te_3$, space group I4/mcm.

The main objective of this work is to study and optimize the temperature dependence of the thermoelectric properties of *Pb* doped Tellurium Telluride $Tl_{8.67}Pb_xSb_{1.33-x}Te_6$ over a wide temperature range and to evaluate the potential of these materials for the thermoelectric applications, e.g. thermoelectric power generator. To develop high-performance bulk thermoelectric materials, we investigated thallium compounds as these types of materials have extremely low thermal conductivity as well as moderate electrical performance [17]. We have investigated the properties of $Tl_{8.67}Pb_xSb_{1.33-x}Te_6$ compound due to their low thermal conductivity and high electrical conductivity and potential for their use as high performance bulk thermoelectric materials.

## II.     Experimental:

The *Pb* doped $Tl_{8.67}Pb_xSb_{1.33-x}Te_6$ (x=0.61, 0.63, 0.65, 0.67, 0.68, 0.70) has been prepared by solid state reactions in evacuated sealed silica tubes. The purpose of this study were mainly for discovering new type of ternary and quaternary compounds by using $Tl^{+1}$, $Pb^{+2}$, $Sn^{+3}$, $Sb^{+3}$ and $Te^{-2}$ elements as the starting materials. Direct synthesis of stoichiometric amount of high purity elements i.e. 99.99 % of different compositions have been prepared for a preliminary investigation. Since most of these starting materials for solid state reactions are sensitive to oxygen and moistures, they were weighing stoichiometric reactants and transferring to the silica tubes in the glove box which is filled with Argon. Then, all constituents were sealed in a quartz tube. Before putting these samples in the resistance furnace for the heating, the silica tubes was put in vacuum line to evacuate the argon and then sealed it. This sealed power were heated up to 650 C° at a rate not exceeding 1 k/mint and kept at that temperature for 24 hours. The sample was cooled down with extremely slow rate to avoid quenching, dislocations, and crystals deformation. The Heating profile for these compounds is: room temperature (RT)-12 h-650 C-24 h-650 C-60 h-560 C-70 h-400C-RT.

Structural analysis of all these samples was carried out by x-rays diffraction, using an Inel powder diffractometer with position-sensitive detector and CuKα radiation at room temperature. No additional peaks were detected in any of the sample discussed here. X-ray powder diffraction patterns confirm the single phase composition of the compounds.

The temperature dependence of Seebeck co-efficient was measured for all these compounds on a cold pressed pellet in rectangular shape, of approximately 5x1x1 mm dimensions. The air sensitivity of theses samples was checked (for one sample) by measuring the thermoelectric power and confirmed that these samples are not sensitive to air. This sample exposes to air more than a week, but no appreciable changes observed in the Seebeck values. The pellet for these measurements was annealed at 400 C for 6 hours.

For the electrical transport measurements 4-probe resistivity method was used and the pellets were cut into rectangular shape with approximate dimension of 5x1x1 mm.

## III.   Results and Discussions:

Structural analysis:

Various concentrations of *Pb* doped $Tl_{8.67}Pb_xSb_{1.33-x}Te_6$ compounds series were synthesized, and their physical properties were studied for x=0.61, 0.63, 0.65, 0.67, 0.68, 0.70. The powder x-rays diffraction pattern at room temperature of the samples is presented in **Fig. 1**. It is found that the tetragonal single phase $Tl_9Sb_1Te_6$ is obtained in the present study. The tetragonal lattice parameters observed at room temperature are *a = 0.866 nm* and *c= 1.305 nm*, in well agreement with the literature data [**18**]. The materials are isostructural with the binary telluride $Tl_5Te_3$, and the crystal structure of $Tl_9Sb_1Te_6$ was determined with the experimental formula, possessing the same space group 14/mcm as $Tl_5Te_3$ and $Tl_8Sb_2Te_6$, in contrast to $Tl_9Sb_1Te_6$ that adopt the space group 14/m.

Physical properties:

To investigate the impact of reduction of the charge carriers in thermal and transport characteristics *Pb* content was increased in $Tl_{8.67}Pb_xSb_{1.33-x}Te_6$ by replacing *Sb* atoms according to the formula. The temperature variation as a function of the Seebeck coefficient (S) for the $Tl_{8.67}Pb_xSb_{1.33-x}Te_6$ (x=0.61, 0.63, 0.65, 0.67, 0.68, 0.70) compounds are shown in **Fig. 3.** The Seebeck coefficient was measured in the temperature gradient of 1 K. The positive Seebeck coefficient increases smoothly with increasing temperature from 300 K to 500 K, for all compounds in particularly for p-type semiconductors having high charge carrier concentration. It is obvious that all the samples exhibit positive Seebeck coefficient for the entire temperature range, indicating that the p-type (hole) carriers conduction dominates the thermoelectric transport in these compounds. When the amount of Pb increased from 0.61 to 0.65, the impact of increasing hole concentration overwhelmed that of the electron scattering, resulting in the reduction of the Seebeck coefficient. On the other hand when the amount of *Pb* doping is larger than 0.65 (as *x* increased further), the *Pb* doping is supposed to increase the carrier's density. However, the smaller grains upon *Pb* doping are believed to be able to enhance the electron scattering, yielding an increase of the Seebeck coefficient and effective mass [**19-20**]. It was found that only an appropriate amount of *Pb* doping could improve the Seebeck coefficient in this particular system. In other words the Seebeck coefficient will drop drastically on doping from the optimum value of *Pb* concentration in this compound. Further improvements could be achieved by (i) employing a rapid fabrication procedure such as melting spinning to reduce the grain size to a much greater degree, (ii) optimizing the doping elements and their corresponding amounts to simultaneously improve the charge mobility and carrier density in order to concurrently enhance the Seebeck coefficient [**21**].
The Seebeck coefficient varies from 80 to 130 $\mu V/K$ as a function of temperature. It can be considered that the lower Fermi energy level caused by decreasing carrier density results in an increased Seebeck coefficient [**22**]. In small x we have the smallest number of charge carriers and large number of holes and thus the largest Seebeck Coefficient is observed. This is indeed the case, as revealed in the Seebeck curves shown in **Fig.4**. In large x i.e. higher *Pb* contents occurs with higher numbers of electrons and thus few charge carriers.
The temperature variations of electrical conductivity of the quaternary compounds are shown in **Fig. 5**. The conductivity observed for all the samples studied here, decreases with increasing temperature, indicating the degenerate semiconductor behavior due to positive temperature coefficient, resulting from the phonons scattering of charge carriers and grains boundaries effects [**23**]. An increasing *x* value, (i.e. increasing the *Pb* deficiency) is expected to increase the number of holes, which is experimental observed. The smaller temperature dependencey may be caused by (less temperature dependence) more grain boundary scattering. No systematic trend was found in the variation of the electrical resistivity for samples $Tl_{8.67}Pb_xSb_{1.33-x}Te_6$ with *Pb* concentration. The low electrical conductivity in the pressureless sintered sample [**23-26**] may be caused by the oxide impurity phase in the grain boundary and the number of the grain boundary. The *Pb* doping level and grain boundary resistance may play important role for increasing electrical conductivity.

The band structure calculations for $Tl_9SbTe_6$ and $Tl_8Sn_2Te_6$ suggest that $Tl_8Sn_2Te_6$ is a heavily doped p-type semiconductor with a partially empty valence band, while the Fermi-level in $Tl_8Sn_2Te_6$ is located in the band gap [5]. In accord with these calculations, both thermal and electrical conductivity measurements indicate that increasing *Pb* contents results in lower values, as determined on sintered polycrystalline samples. For the Seebeck coefficient, we have observed an opposite trend; increasing *Pb* concentration, causes Seebeck values increases.
The temperature behavior as the doping concentration *x*, is varied, and can be attributed to the relationship between Seebeck coefficient, temperature and injected charge carrier concentration displayed below [10]:

$$s = \frac{8\pi^2 k_B^2}{3eh^2} m^* . T (\frac{\pi}{3n})^{2/3}$$

Where, $K_B$, is the Boltzmann constant, *e* is the electronic charge, *h* is the Plank's constant, $m^*$ is the effective mass, and *n* is the charge carrier concentration. The carrier concentration and effective mass are two major factors affecting the Seebeck coefficient. Hence the samples with high, and thus temperature independent *n* (small *x*), will exhibit linearly increasing Seebeck coefficient with temperature. Those with small *n* and large x, that increases with temperature, causes a decrease in the Seebeck coefficient, once sufficient charge carriers are able to cross the band gap of the compound studied. Using $E_{gap}=2eS_{max}T_{max}$ [24], a gap of 0.23 eV has been calculated for $Tl_{8.67}Pb_{0.70}Sb_{0.63}Te_6$ system.

The electrical conductivity, σ, for the $Tl_{8.67}Pb_xSb_{1.33-x}Te_6$ compounds with $0.61 \leq x \leq 0.70$ as shown in **Fig. 5**, decreases with increase in temperature across the entire temperature range examined; these results are the indicative of metallic behavior, an evidence of a relatively high carrier concentration. Increased, doping concentration, causes decrease in σ, as expected and inversely affecting their Seebeck counterpart. The $Tl_{8.67}Pb_{0.70}Sb_{0.60}Te_6$ displays the highest value of 1650 $(\Omega\text{-cm})^{-1}$ at 290 K, and $Tl_{8.67}Pb_{0.65}Sb_{0.65}Te_6$ displays the lowest with 545 $(\Omega\text{-cm})^{-1}$ and the samples with x= 0.61 and 0.63, almost have very close values of about 850 and 900 $(\Omega\text{-cm})^{-1}$ respectively. The conductivity differences at room temperature between the hot pressed pellet and the ingot was observed a little change from 850 to 855 $(\Omega\text{-cm})^{-1}$ respectively for the $Tl_{8.67}Pb_{0.63}Sb_{0.70}Te_6$ compound.
To enhance the power factor (PF= $S^2$ σ) for these compounds, we need to decouple the electrical conductivity from the Seebeck co-efficient, usually inversely proportional to each other in these systems. The main contribution in the PF comes from the Seebeck co-efficient, so we have to design the materials such that their *S* should be enhanced. The power factors calculated from the electrical conductivity σ, and the Seebeck co-efficient S, obtained for these compounds are displayed in **Fig. 6**. The power factor increases with increasing temperature for all these compounds. The power factor shows very complex behavior with the doping of Pb concentrations. The $Tl_{8.67}Pb_{0.60}Sb_{0.70}Te_6$ compound display the highest value 8.56 ($\mu Wtt\text{-}cm^{-1}\text{-}K^{-2}$) of PF at 550 K and 3.52 56 ($\mu Wtt\text{-}cm^{-1}\text{-}K^{-2}$) at 290 K. The lowest PF factors were observed for $Tl_{8.67}Pb_{0.70}Sb_{0.60}Te_6$ compound which have values of 1.66 56 ($\mu Wtt\text{-}cm^{-1}\text{-}K^{-2}$) at 550 K and 1.01 56 ($\mu Wtt\text{-}cm^{-1}\text{-}K^{-2}$) at 290 K. As discussed before, an increasing the *Pb* deficiency, is expected to increase the number of holes, the dominant charge carriers. This expected trend is experimental observed: with increasing x, the Seebeck co-efficient decreases and then increases after optimized value of Pb concentration, which results in increase in the S. The smaller temperature dependence $Tl_{8.67}Pb_{0.70}Sb_{0.60}Te_6$ compound may be caused by (less temperature dependence) more grain boundary scattering.

## IV.     Conclusion:

Various concentrations of *Pb* doped $Tl_{8.67}Pb_xSb_{1.33-x}Te_6$ compounds series were synthesized, and their physical properties were studied for x=0.61, 0.63, 0.65, 0.67, 0.68, 0.70. These materials are isostructural with the binary telluride $Tl_5Te_3$, and the crystal structure of $Tl_8Sb_2Te_6$ was determined with the experimental formula, possessing the same space group 14/mcm as $Tl_5Te_3$, where as in contrast to $Tl_9Sb_1Te_6$ that adopt the space group 14/m.

The thermoelectric properties of polycrystalline $Tl_{8.67}Pb_xSb_{1.33-x}Te_6$ compounds were measured in the temperature range from room temperature to about 550 K. The Seebeck coefficient is positive in the whole temperature range, and is increasing with increase in temperature, conferring that hole conduction dominates in these compounds. For higher trends of Pb the increasing hole concentration overwhelmed that of the electron scattering, resulting in the reduction of the Seebeck coefficient. On the other hand when the amount of Pb doping is more than 0.65, the Pb doping is supposed to increase the carrier's density. However, the smaller grains upon Pb doping are believed to enhance the electron scattering, yielding an increase of the Seebeck coefficient. Further improvements appear to be possible by optimizing the materials on both the micro and nano level, as demonstrated for PbTe in different studies. Moreover, different partial substitutions may enhance, or decrease the performance, such as introducing lanthanoids or Sn atoms, which remains to be investigated. This work is a best example of optimizing dopants concentration to achieve desirable thermoelectric properties in *Pb* doped $Tl_{8.67}Pb_xSb_{1.33-x}Te_6$ compounds

**Figure Captions:**

Figure 1.: XRD data of $Tl_{8.67}Pb_xSb_{1.33-x}Te_6$ (x=0.61, 0.63, 0.65, 0.67, 0.68, 0.70), collected at room temperature.

Figure 2.:
EDX data for $Tl_{8.67}Pb_{0.63}Sb_{0.70}Te_6$ , collected at room temperature are shown here, to conform the elemental analysis in the system, with all other compounds have the same stoichiometric ratio as designed.

Figure 3.:
Temperature dependence of Seebeck Co-efficient of $Tl_{8.67}Pb_xSb_{1.33-x}Te_6$ (x=0.61, 0.63, 0.65, 0.67, 0.68, 0.70).

Figure 4.:
Seebeck co-efficient dependence on $x$ ($x$=0.61, 0.63, 0.65, 0.67, 0.68, 0.70) are shown for $Tl_{8.67}Pb_xSb_{1.33-x}Te_6$ at 305 and 550 K.

Figure 5.:
Temperature dependence of conductivity of $Tl_{8.67}Pb_xSb_{1.33-x}Te_6$ (x=0.61, 0.63, 0.65, 0.67, 0.68, 0.70) of the cooled pressed pellet, with heating profile.

Figure 6.:
Variation of Power factor with temperature and their dependency on doping concentration for $Tl_{8.67}Pb_xSb_{1.33-x}Te_6$ (x=0.61, 0.63, 0.65, 0.67, 0.68, 0.70).

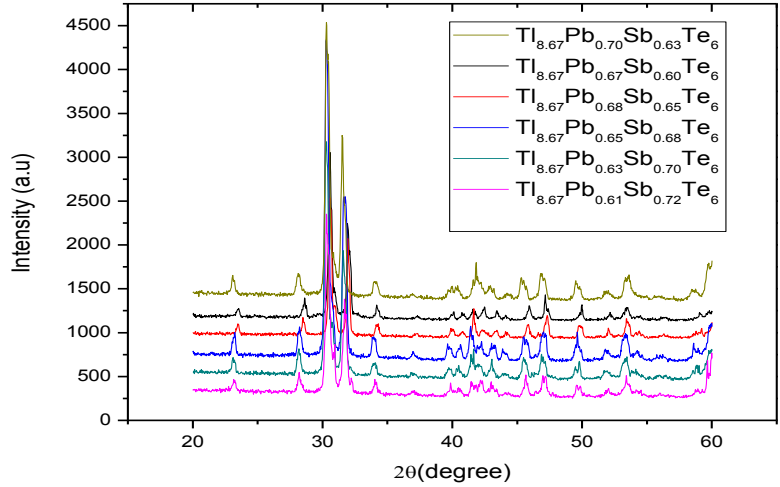

Figure 1.:
XRD data of $Tl_{8.67}Pb_xSb_{1.33-x}Te_6$ (x=0.61, 0.63, 0.65, 0.67, 0.68, 0.70), collected at room temperature.

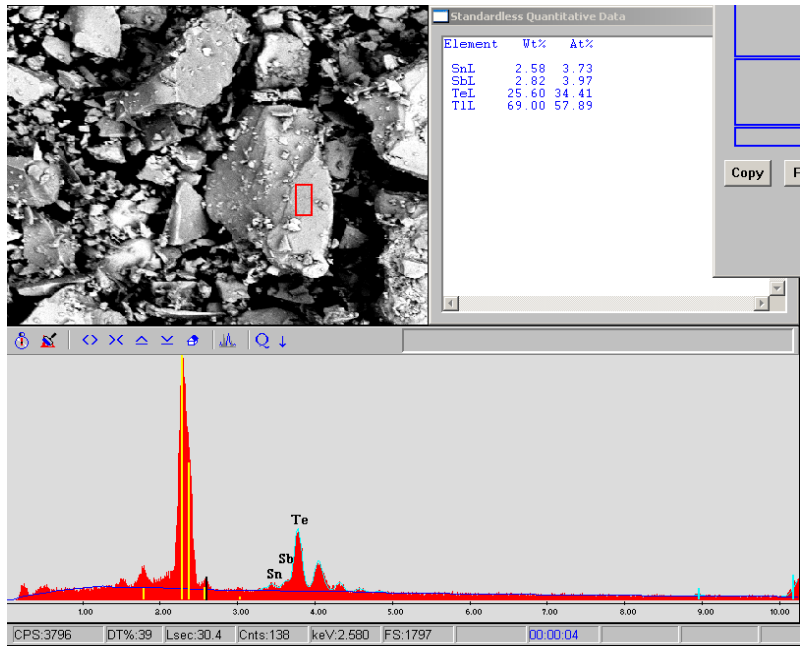

Figure 2.: EDX data for $Tl_{8.67}Pb_{0.63}Sb_{0.70}Te_6$, collected at room temperature are shown here, to conform the elemental analysis in the system, with all other compounds have the same stoichiometric ratio as designed.

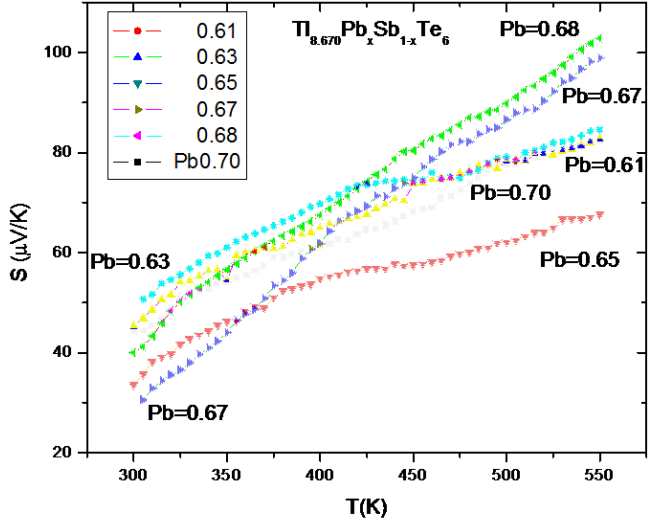

Figure 3.: Temperature dependence of Seebeck Co-efficient of $Tl_{8.67}Pb_xSb_{1.33-x}Te_6$ (x=0.61, 0.63, 0.65, 0.67, 0.68, 0.70).

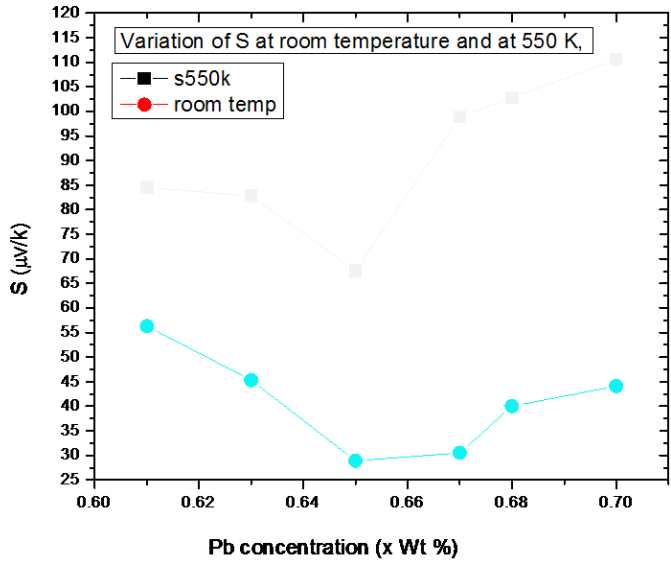

Figure 4.: Seebeck co-efficient dependence on *x* (*x*=0.61, 0.63, 0.65, 0.67, 0.68, 0.70) are shown for $Tl_{8.67}Pb_xSb_{1.33-x}Te_6$ at 305 and 550 K.

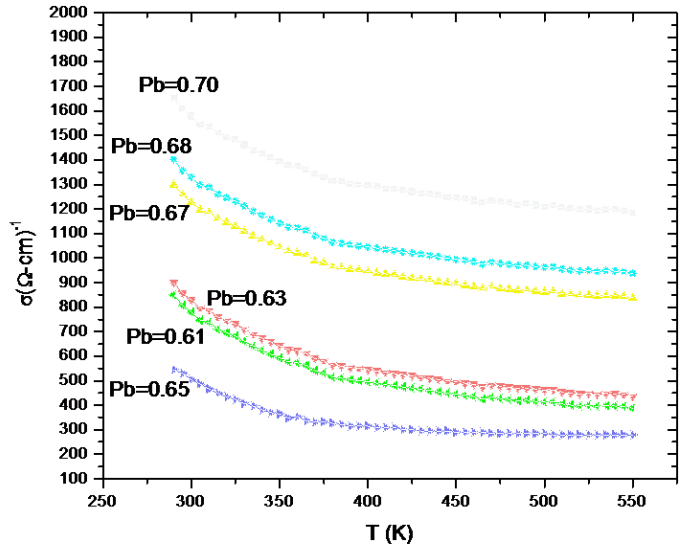

Figure 5.: Temperature dependence of conductivity of $Tl_{8.67}Pb_xSb_{1.33-x}Te_6$ (x=0.61, 0.63, 0.65, 0.67, 0.68, 0.70) of the cooled pressed pellet, with heating profile.

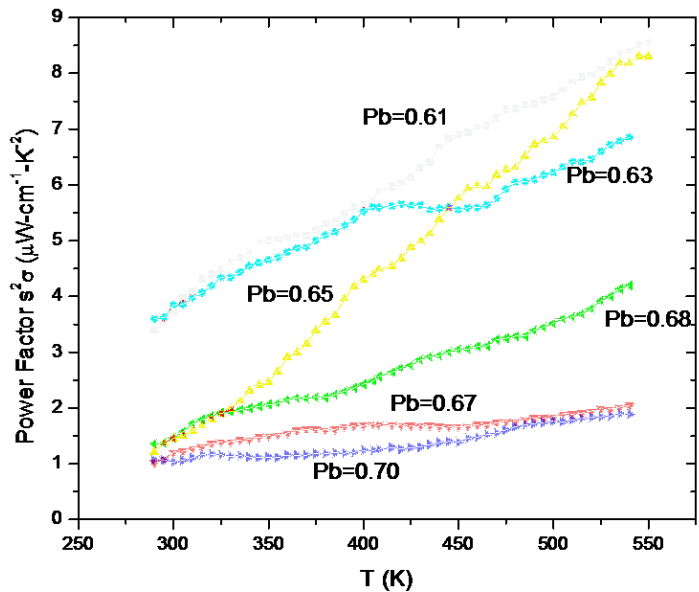

Figure 6.: Variation of Power factor with temperature and their dependency on doping concentration for $Tl_{8.67}Pb_xSb_{1.33-x}Te_6$ (x=0.61, 0.63, 0.65, 0.67, 0.68, 0.70).